# Emergent many-body composite excitations of interacting spin-1/2 trimers

Anup Kumar Bera[1,2], S. M. Yusuf[1,2*], Sudip Kumar Saha[3], Manoranjan Kumar[3], David Voneshen[4,5], Yurii Skourski[6] and Sergei A. Zvyagin[6]

[1]*Solid State Physics Division, Bhabha Atomic Research Centre, Mumbai 40085, India*
[2]*Homi Bhabha National Institute, Anushaktinagar, Mumbai 400094, India*
[3]*S. N. Bose National Centre for Basic Sciences, Block JD, Sector III, Salt Lake, Kolkata 700106, India*
[4]*ISIS Facility, STFC Rutherford Appleton Laboratory, Harwell Oxford, Didcot OX11 0QX, UK*
[5]*Department of Physics, Royal Holloway University of London, Egham, TW20 0EX, UK*
[6]*Dresden High Magnetic Field Laboratory (HLD-EMFL), Helmholtz-Zentrum Dresden-Rossendorf, 01328 Dresden, Germany*

*e-mail: smyusuf@barc.gov.in

**Understanding exotic forms of magnetism in quantum spin systems is an emergent topic of modern condensed matter physics. Quantum dynamics can be described by particle-like carriers of information, known-as quasiparticles that appear from the collective behaviour of the underlying system. Spinon excitations, governing the excitations of quantum spin-systems, have been accurately calculated and precisely verified experimentally for the antiferromagnetic chain model. However, identification and characterization of novel quasiparticles emerging from the topological excitations of the spin system having periodic exchange interactions are yet to be obtained. Here, we report the identification of emergent composite excitations of the novel quasiparticles doublons and quartons in spin-1/2 trimer-chain antiferromagnet $Na_2Cu_3Ge_4O_{12}$ (having periodic intrachain exchange interactions $J_1$-$J_1$-$J_2$) and its topologically protected quantum 1/3 magnetization-plateau state. The characteristic energies, dispersion relations, and dynamical structure factor of neutron scattering as well as macroscopic quantum 1/3 magnetization-plateau state are in good agreement with the state-of-the-art dynamical density matrix renormalization group calculations.**

The magnetism in one-dimension (1D), dominated by quantum fluctuations, remains one of the extraordinary fundamental topics over the last century and has been explored extensively since the early days of quantum mechanics. In 1931, Bethe stated an ansatz to predict the ground state of 1D spin-1/2 Heisenberg antiferromagnetic chain (HAC) system[1], and further extensions of this ansatz[2-5] successfully predict the ground state as well as the excitations of the many-body interactions of the generalized 1D anisotropic Heisenberg–Ising (or XXZ) antiferromagnetic (AFM) model. Important predictions, such as the spinon excitations with fractional spin-1/2 quantum number, characteristic spinon continuum, as well as many body Bethe string states have subsequently been accurately calculated and precisely verified experimentally[6-11]. However, the utility of the Bethe ansatz is limited when perturbing the HAC beyond the solvable uniform XXZ model[12,13], such as, with the periodic exchange interactions within the chain. One of such models is a quantum trimer spin-chain with repeating couplings $J_1$-$J_1$-$J_2$ (intratrimer $J_1$, and intertrimer $J_2$)[14-19] for which emergent composite excitations of novel quasi-particles doublons and quartons, in addition to the low energy fractional spinon excitations, have been predicted very recently[20]. Such emergent composite excitations appear only over a specific range of small values of $\alpha=J_2/J_1$ <0.4, and with increasing $\alpha$ value those states fractionalize to form the conventional spinon continuum when $\alpha \to 1$. However, a real material which meets these stringent requirements to support these composite excitations has thus far remained elusive. Here, we perform inelastic neutron scattering (INS) measurements on $Na_2Cu_3Ge_4O_{12}$ and report the experimental realization of the emergent composite excitations of the novel quasi-particles, doublons and quartons. The INS spectra reveal that the doublon and quarton excitations govern the high energy dynamics, whereas spinon excitations are dominant at low energies. Our experimentally obtained eigenstates, dispersion relations, and dynamical structure factor are compared in detail to the state-of-the-art dynamical density matrix renormalization group (DMRG) calculations of the 1D spin-1/2 trimer model. The excellent agreement between the experiment and theory allows us the unambiguous identification and full characterization of the emergent composite excitations. Such emergent states are characterized by strong quantum entanglement, having promising prospective for quantum information processing.[21]

## Results

**Spin Hamiltonian and magnetization plateau.** $Na_2Cu_3Ge_4O_{12}$ is an excellent realization of the paradigmatic spin model of spin-1/2 HAC comprising of coupled spin-trimers[22,23]. The crystal structure of $Na_2Cu_3Ge_4O_{12}$ is composed of $Cu_3O_8$ trimers formed by three edge-sharing $CuO_4$ square planes in a linear fashion [Figs. 1a, 1b, and Supplementary Note 1]. The magnetic $Cu^{2+}$ ions within the CuO$_4$ square planes have quantum spin-1/2 ($S$=1/2) providing a close approximation to the spin Hamiltonian

$$H = \sum_{r=1}^{N/3}[J_1(\vec{S}_{r,1}\cdot\vec{S}_{r,2} + \vec{S}_{r,2}\cdot\vec{S}_{r,3}) + J_2(\vec{S}_{r,3}\cdot\vec{S}_{r+1,1}) + J_3(\vec{S}_{r,1}\cdot\vec{S}_{r,3})] \qquad (1)$$

where $\vec{S}_{r,i}$ is a spin-1/2 operator at $i^{th}$ (=1,2, and 3) site of $r^{th}$ spin-trimer. The positive $J$s denote the dominating antiferromagnetic exchange interactions with the nearest-neighbour (NN) superexchange interaction $J_1$ within a given trimer via intermediated oxygen ions, intertrimer super-superexchange interaction $J_2$ (= $\alpha J_1$) via oxygen, germanium, and oxygen ions. The $J_3$ (= $\beta J_1$) denotes the next-nearest-neighbour (NNN) intratrimer super-superexchange interaction between two edge spins of a given trimer [Fig. 1b] via oxygen-oxygen ions. For $\alpha$=0, the trimers are isolated from each other, while for $\alpha$=1 and $J_3$ = 0 the model reduces to the isotropic Bethe-ansatz soluble HAC[1]. The antiferromagnetic interaction $J_3$ competes with the interaction $J_1$ and introduces a frustration in the spin system. In zero field, Na$_2$Cu$_3$Ge$_4$O$_{12}$ undergoes a magnetic phase transition to a long-range Néel type antiferromagnetic ordered state below the $T_N \sim 2$ K[22] due to a weak interchain interaction $J_4$ (= $\gamma J_1$) [not included in equation (1)], through the super-super exchange pathways via the nonmagnetic Na$^+$ and Ge$^{4+}$ ions. Such weak interaction $J_4$ is a small perturbation to the model Hamiltonian in equation (1) and slightly modifies the low energy excitations (discussed later). Nevertheless, above the $T_N$, Na$_2$Cu$_3$Ge$_4$O$_{12}$ retains the main characteristics of the HAC comprising of coupled spin-trimers and well approximated by the 1D HAC model [equation (1)]. The experimentally measured temperature dependent susceptibility [$\chi(T)$] and the pulse-field magnetization [$M(H)$] curves are well reproduced by the model Hamiltonian in equation (1) [Fig. 1c-d] with the parameters $J_1$ = 235 K, $\alpha$ = 0.18, and $\beta$ = 0.18 (for details see Methods and supplementary Fig. 6) and Landé $g$-factor $g$ =2.06 (as determined by ESR study; see Supplementary Note 3 for details). The presence of a 1/3 magnetization plateau, a characteristic feature of the weakly-coupled trimer spin systems, is found above $\mu_0 H_{C1}$ = 28 Tesla (T) both experimentally and theoretically [Fig. 1d]. Here, the 1/3-magnetization plateau is a macroscopic phenomenon of quantum origin. The plateau state is driven by the tendency of neighbouring antiferromagnetically coupled spins to form highly entangled spin-singlet states, leading to the spontaneously broken translational symmetry[15,24,25]. This plateau follows the Oshikawa-Yamanaka-Affleck rule[24] $S \cdot p \, (1 - M/M_{sat}) = Z$ where $S$ is the spin of the system, $p$ is the number of spins per unit cell, $M$ is the magnetization measured in the unit of saturation magnetization $M_{sat}$, and $Z$ is a set of integer numbers. Our precise DMRG calculations (system sizes up to $N$=96) reveal that the 1/3 magnetization plateau state persists up to a magnetic field of $\mu_0H_{C2}$=252.5 T and the field polarized state appears above $\mu_0H_S$~ 265.8 T. The topologically nontrivial 1/3 magnetization plateau state can be characterized by a nonzero integer Chern number and appears due to the existence of finite gaps in the thermodynamic limit with the width of plateaus proportional to the gap size[17]. The topological origin of such phenomena of quantized magnetization plateaus relates the plateau state to a correlated topological insulator[17].

**Energy level spectrum.** The schematic energy level spectrum and corresponding Eigen functions of coupled spin-1/2 trimers are shown in Fig. 2a. $Na_2Cu_3Ge_4O_{12}$ resembles weakly coupled trimers and its spectrum is mostly dominated by the isolated trimer. For an isolated spin-1/2 trimer ($\alpha=\gamma=0$) with the exchange interactions $J_1$, the ground state of a trimer system is a doublet with two fold degenerate lowest energy $E_0= -J_1$ for $\beta$=0 and $E_0$=-0.955$J_1$ for $\beta$=0.18. The wave-function is a linear combination of one singlet and a free spin where singlet bond is always between the NN bonds (Fig. 2a). The doublet lowest excited state is also doubly degenerate and is at energy $E_1$=0 for $\beta$=0 and $E_1$=-0.135$J_1$ for $\beta$=0.18. Their wave function is a product of free spin with a singlet bond between NNN spins (Fig. 2a). The highest excitation is a quartet with total spin = 3/2; these energies are four-fold degenerate and are located at $E_2$=0.5$J_1$ for $\beta$ = 0 and $E_2$ = 0.545$J_1$ for $\beta$=0.18. The wave function is a linear combination of configurations which are direct product of a free spin and a triplet bond between two NN spins and product of a free spin and triplet bond between two NNN spins. All three spins in the same direction is another configuration for quartet (Fig 2a). The weak antiferromagnetic exchange interaction between neighbouring trimers (smaller values of $\alpha$) gives rise to a singlet ground state, since the lowest energy state of each trimer behaves as effective spin-1/2. A possible configuration of antiferromagnetically coupled trimers is shown in Fig. 2b. The lowest excitation energy modes are spinon modes generated by flipping one spin on trimer and it disperses throughout the system. One of the configurations of this excitation is shown in Fig. 2b. Higher energy excitation modes can be created by the flipping of a down spin in the initial configuration as well as a singlet bond between NN spins replaced by a singlet bond with NNN spins. This type of dispersing excitation is called a doublon (the terms doublon and quarton used here were introduced in Ref.[26]) where the change in total $S^z$ is still one, as for the case in Fig. 2b. Another type of dispersive excitation is quarton where a single spin is flipped on a trimer to break the singlet bonds and polarize all three spins on a trimer or to convert the initial trimer configuration to a product of a free spin and a triplet bond between NNN (or NN) spins (Fig. 2b). The horizontal arrow in Fig. 2b represents the delocalization of an excitation. The doublon and quarton excitations, originating from internal trimer states, are almost localized for small values of $\alpha$, and cannot be classified as standard magnons, or triplons, or spinons. For the larger values of $\alpha$, the doublons and quartons lose their identity and fractionalize into the standard spinon continuum that emerges for $\alpha \to 1$, *i.e.* for the HAC. On the other hand, in the intermediate regime, doublon and quarton coexist with spinon-pair continua. The theoretical dispersion relations and the intensities of the spinon, doublon, and quarton excitations calculated by the hybrid ED/DMRG (see Method) for $J_1 = 235$ K, $\alpha = 0.18$, $\beta = 0.18$ and $g = 2.06$ which correspond to $Na_2Cu_3Ge_4O_{12}$ are illustrated in Fig. 2c. Our dynamical results agree well with the results of several other numerical methods [ED[26], ED with truncated Hilbert space[20], Quantum Monte Carlo by applying a variant of the stochastic analytic continuation (QMC-SAC)[20]] (see Supplementary Note 6 for further details). The composite doublon, and quarton excitation states (marked as B and C, respectively) are gapped throughout reciprocal space, whereas, spinon-pair continua (marked as A) are gapless at specific wave vectors. All the three excitation modes have distinct energy ranges where the spinon-pair modes are found below 5 meV, while, the doublon, and quarton states appear at intermediate ($E = \hbar\omega$ over 17-22 meV) and high (32-37 meV) energy ranges, respectively. The broadening of the doublon and quarton excitations appear due to their weak dispersion relations governed by the weaker $J_2$ coupling ($\alpha = 0.18$)[26]. For spinons, most of the intensity appears around the zone centre ($q = \pi$). The spectral weight of quarton modes are concentrated over $q = (0.5\text{-}1.5)\pi$. Whereas, spectral weight of doublon modes are distributed over $q = (0.3\text{-}0.8)\pi$ and $(1.2\text{-}1.7)\pi$ (Fig. 2c).

**Spin Excitation.** We performed INS experiments to measure the dynamical structure factors $S(Q, \omega)$ of $Na_2Cu_3Ge_4O_{12}$ (see Methods) in the 1D state at 3 K ($T_N = 2$ K). Figures 3a and 3c depict the phonon background corrected magnetic excitation spectra of $Na_2Cu_3Ge_4O_{12}$ powders, measured with incident energies $E_i = 45$ and 9.6 meV, as a function of energy and wave vector which cover all the three spin excitation levels at ~ 3, 16, and 30 meV (A, B, and C). Although the directional-dependent information in the $S(Q, \omega)$ is lost due to the powder averaging, the powder $S(|Q|, \omega)$ preserves singularities arising in the density of states as a function of $E = \hbar\omega$ and provides distinctive fingerprints of the Hamiltonian (1) which can be readily compared to theoretical calculations to estimate the $J$ parameters. The higher energy modes B and C are gapped and weakly dispersive. Whereas, the low energy mode A is clearly dispersive in nature and shows a possible gapless minima at around wave vector transfer $|Q| = 1.2$ Å$^{-1}$. The experimental spectra for $Na_2Cu_3Ge_4O_{12}$ are in good agreement with the powder averaged DMRG results of DSF of a coupled trimer HAC system [Equation (1)] with the parameters $J_1 = 235$ K, $\alpha = 0.18$, and $\beta = 0.18$ (Figs. 3b,d) in terms of the Eigen energies as well as spectral features. Such an agreement thus unambiguously provides an experimental realization of doublon and quarton excitations and their in-depth characterization. An energy cut through the data (integration of $|Q|$ over 0-3 Å$^{-1}$) (Fig. 3e) reveals all the three excitation modes as well as their respective intensity and spectral width. Comparisons with the theoretical DMRG results (shown by the red curves) have strong agreements overall, while the slight difference at low energies could be ascribed to the effects of 3D interchain couplings, which are weak and have not been considered in the calculations. Further, the temperature dependent neutron scattering spectra (Fig. 3f,3g) reveal that the doublon and quarton excitations persist up to ~ 250 K, which is within the energy range of the intra-trimer NN exchange constant $J_1 = 235$ K, suggesting internal trimer excitations in the coupled trimer chain as their origin.

**Phase diagram.** The effect of inter-trimer exchange coupling is illustrated by the simulated quantum phase diagram of the 1D trimer model [Equation (1)] in the extended $\alpha$-$H$ space for $\beta=0.18$ and $J_1 = 235$ K [Fig. 4]. To explore the $\alpha$-$H$ space experimentally, the value of $\alpha$ can be tuned by a suitable chemical substitution or an application of pressure. For $\alpha=0$, the model is reduced to isolated trimers. In zero magnetic field, such isolated trimer system has discrete energy levels, with a doublet as the ground state. For an isolated trimer system, an application of an infinitesimal magnetic field leads to the onset of the 1/3 magnetization plateau state. The plateau phase exhibits the stability of the doublet state of the trimer. On increase in the applied magnetic field, the system shows a transition from the doublet state $M=1/2$ (magnetization plateau state) to the quartet state $M=3/2$ (the saturation magnetization state), and the magnetic field required for the transition is proportional to the plateau width $H_{c2}$. The condition for doublet to quartet transition is $E_0 - \frac{1}{2}g\mu_B H_{c2} = E_2 - \frac{3}{2}g\mu_B H_{c2}$ where $E_0$ and $E_2$ are the energy values for the doublet ground state and the quartet excited states in absence of magnetic field, respectively. The width of the 1/3 plateau is proportional to $(E_2 - E_0)$ $(= 1.5\ J_1)$. Now we turn our attention to an interacting trimer spin system as applicable for the present compound. In such a situation, any finite intertrimer interaction $J_2$ induces a dispersive band for the doublet and quartet energy states. In this case also the 1/3 magnetization state gets stabilized, however, it requires a finite magnetic field value $H_{c1}$ to induce an onset of the 1/3 magnetization plateau state. The $H_{c1}$ increases with the increasing $J_2$, whereas, the $H_{c2}$ (the field at which the 1/3 magnetization plateau state ends) reduces with increasing $J_2$. The width of plateau ($H_{c2}$ -$H_{c1}$) thus decreases with increasing $J_2$. The intertrimer coupling $J_2$ also gives another threshold beyond the $H_{c2}$ to reach the magnetization to its complete saturation. Such a state between the $H_{c2}$ and the saturation magnetic field $H_S$ is denoted as a meta-magnetic state [Fig. 4]. On the other hand, below the 1/3 plateau (i.e. $H < H_{c1}$) all the magnetic states are denoted as low magnetic states [Fig. 4]. In the low magnetic states, the *M-H* curve varies linearly with *H* in the lower magnetization regime, but it has an algebraic variation as $M(H)-M(H_{c1}) \propto (H-H_{c1})^{1/2}$ near the cusp of the plateau at $H_{c1}$. The coefficient ½ of variation is reported near the cusp for various model quantum spin systems[27,28]. The phase diagram with the frustrated interaction $J_3$ in the ($\beta$-$J_1$) plane is shown in Supplementary Note 5.

Within the topologically protected quantum 1/3 magnetization-plateau state of the trimer chain, the nearest-neighbour spin-spin correlations and bipartite entanglement on three bonds ($S_{3i-2,3i-1}$; $S_{3i-1,3i}$, and $S_{3i,3i-1}$) also exhibit interesting correlation[29]. Finite value of the bipartite quantum correlations in the trimer ground state, that persists up to a very high temperature (asymptotic limit of temperature $T \to \infty$) has a potential role in quantum information processing[21]. Here the possible emergence of novel quasi-particles, by reconstruction of elementary excitations due to spontaneous symmetry breaking, is expected up to a temperature close to room temperature. Further, the quantum entanglement for $Na_2Cu_3Ge_4O_{12}$ reveals a very high decoherence (critical) temperatures of ~ 250 K (the quantum entanglement estimated following the Hilbert-Schmidt norm[18] for $\alpha = 0$ (see Supplementary Note 4, and Supplementary Fig. 5 for details). Such a high decoherence temperature, close to room temperature, is of special requirement for practical device applications.

In summary, the central finding of the present study is that $Na_2Cu_3Ge_4O_{12}$ is the first compound to realize the weakly coupled AFM trimer chain model and to reveal experimental observation of topological quantum excitations of novel quasi particles doublon and quarton. The present work, thus, is expected to open up an avenue for the exploration of novel topological quantum states based on spin-trimers systems and beyond. The present work anticipates new directions of future theoretical and experimental studies to explore (i) high energy fractionalization mechanism, (ii) synthesis of new materials or modifying the chemical composition to achieve low-field magnetization plateau state, as well as to vary the excitation energies and gaps of the newly observed topological quantum excitations, (iii) the evolution of the different types of excitations, especially when the doublons and quartons begin to form the conventional spinon continuum with the increasing inter-trimer exchange interaction ($\alpha = J_2/J_1$), and (iv) the quantum topological excitation states at higher temperatures extending up to room temperature or beyond.

**Methods:**

**Sample preparation and characterization.** Polycrystalline samples of $Na_2Cu_3Ge_4O_{12}$ were prepared through the solid state reaction method. High purity (>99.99%) reagents of $Na_2CO_3$, CuO, and $GeO_2$ were mixed together with a molar ratio 1:3:4. The mixture was annealed at 800 °C for total 40 h in air in a muffle furnace with several intermediate grindings. The phase purity of the powder samples was confirmed by Rietveld analysis of the room temperature x-ray diffraction pattern measured using a laboratory x-ray machine (see Supplementary Fig. 1).

**Neutron diffraction.** Room temperature neutron powder diffraction pattern was measured on the powder diffractometer PD-II ($\lambda$ = 1.2443 Å) at Dhruva reactor, Bhabha Atomic Research Centre, India (to derive crystal structural correlations) (see Supplementary Fig. 2).

**Dc-magnetization using PPMS.** The temperature and field dependence of magnetization were measured using a commercial physical properties measurement system (PPMS) (Cryogenic Co. Ltd., UK). The dc-magnetization measurements were carried out over 1.5-300 K in the zero-field-cooled condition under 1 T of magnetic fields. Isothermal $M$ vs $H$ curve was measured at 3.0 K over the field range of ± 9 T (see Supplementary Note 2).

**High-pulse field magnetization.** The high pulse-field magnetization measurements up to 60 T were performed at the Hochfeld Magnetlabor, HZDR, Dresden, Germany. The magnetization signal was detected by an induction method[30] and corrected for the empty magnetometer background to obtain the sample magnetization.

**Inelastic neutron scattering.** The inelastic neutron scattering (INS) measurements were performed on the high-flux neutron time-of-flight instruments MERLIN at the ISIS facility of the Rutherford Appleton Laboratory, Didcot, United Kingdom[31]. The INS spectra on MERLIN were recorded with a fixed incident neutron energy of $E_i$ = 18 meV with the repetition rate multiplication (RRM) method[32,33] by using a straight Gd Fermi chopper (speed was fixed to 250 Hz), which provides simultaneous measurements of INS patterns corresponding to incident energies of $E_i$ = 45, 18, and 9.6 meV, respectively. About a 17-g powder sample was used for these INS measurements. The INS spectra were collected at several temperatures down to 3 K using a helium cryostat. The INS data were reduced using the MANTID software package[34]. The raw data were corrected for detector efficiency and time-independent background following standard procedures. The INS spectra are corrected for the background due to the phonon scatterings. The energy and wave vector dependent phonon backgrounds are estimated from the measured spectra at 300 K with an application of a required Bose factor.

**High-field ESR.** The electron spin resonance (ESR) measurements were performed employing a 16 T transmission-type ESR spectrometer[35]. Measurements were done in a frequency of 136 GHz at 7.3 and 18.7 K, using a VDI microwave-chain radiation source (product of Virginia Diodes, Inc., USA). An InSb hot-electron bolometer (QMC Instruments Ltd., UK) was used to record the spectra.

**Numerical calculations.** The temperature dependence of the thermodynamic quantities for $Na_2Cu_3Ge_4O_{12}$ were calculated using the hybrid exact diagonalization (ED) and density matrix renormalization group (DMRG) method (hybrid ED/DMRG)[36,37]. It combines exact diagonalization (ED) of short chains with density matrix renormalization group (DMRG) calculations of progressively longer chains. The hybrid ED/DMRG technique has been extensively applied to 1D spin systems[36] and fermionic systems[37]. The main advantage of this method is that the full spectrum of large systems of N spins is not needed. Since thermal fluctuations limit the range of spin correlations, small systems using ED provides the thermodynamic limit at high temperature where the system size exceeds the correlation length. However, the study of thermodynamics at low temperature requires systems of large number of spins to reach the thermodynamic limit due to presence of large correlation length. The partition function is the sum of the Boltzmann probability of all energy states, and has a significant contribution from low energy states at low temperature. The higher excited states have exponentially small contributions, therefore, the accurate low temperature properties require only accurate low-lying energy states. The DMRG method is well known for its accurate calculation of low-lying states and this property of DMRG is exploited to calculate the low temperature properties.

In the hybrid ED/DMRG method density matrix of the system block is calculated using the projection of all low lying energy states of the superblock and the details of projection procedure is explained in ref. [36]. For the trimer model in Eq. 1, we have used ED to calculate the thermodynamics at high temperature and progressively larger system size using DMRG yields low energy excitations and extends the thermodynamics to lower temperature. We have retained up to 700 eigenstates in the density matrix for the DMRG calculation and calculated 400 states in each $S^Z$ sector. It gives access to the accurate temperature dependence down to $T \sim 0.01\ J_1$. The Eqn. (1) was used with different sets of values of $J_1$, $\alpha$, and $\beta$ to reproduce the experimentally observed $\chi(T)$ and $M(H)$.

The dynamical spin structure factor for $Na_2Cu_3Ge_4O_{12}$ was calculated using the DMRG method along with the correction vector method[38-40] for $N = 48$ spins i.e., 16 trimers (Figs. 2c and 3b). To compare the calculated spinon excitation spectrum with the high resolution experimental INS spectrum with $E_i = 9.6$ meV (Fig. 3c), the low energy spinon spectrum is also calculated for $N = 96$ spins i.e., 32 trimers (Fig. 3d) which minimizes the finite size effect. The dynamical structure factor is defined as

$$S(q,\omega) = \sum_n \frac{|\langle \psi_n | S_q^\alpha | \psi_0 \rangle|^2}{E_n - (E_0 + \omega) + i\eta} \quad (1)$$

Here $E_0$ and $E_n$ are the energies of the ground state and $n^{th}$ excited state, respectively. The $\omega$ and $q$ represent the energy and momentum, respectively, transferred to the lattice. The $\eta$ is the broadening factor. $|\psi_0\rangle$ is the ground state wave function and $|\psi_n\rangle$ is the $n^{th}$ excited state wavefunction. If $a$ denotes the $x$, $y$, and $z$ component of spin, we can define

$$S_q^\alpha = \sqrt{\frac{2\pi}{N}} \sum_i e^{iqj} S_j^\alpha \quad (2)$$

**Data availability:**

The datasets for the inelastic neutron scattering experiment on the time-of flight MERLIN spectrometer are available from the ISIS facility, Rutherford Appleton Laboratory data portal (DOI: 10.5286/ISIS.E.RB1910432).

**Code availability:**

The code is available upon reasonable request from M. K.

**Acknowledgements:** A.K.B. and S.M.Y. thank the Department of Science and Technology, India (SR/NM/Z-07/2015) for financial support to carry out the neutron scattering experiment, and Jawaharlal Nehru Centre for Advanced Scientific Research (JNCASR) for managing the project. M.K. thanks SERB for financial support through Grant Sanction No. CRG/2020/000754. This work was supported by the Deutsche Forschungsgemeinschaft (through ZV 6/2-2) and by HLD at HZDR, member of the European Magnetic Field Laboratory (EMFL).

**Author contribution:**

A.K.B. and S.M.Y. conceived the project. A.K.B. prepared and characterized the samples. A.K.B. and D.V. performed the neutron scattering experiments. S.A.Z. and Y.S. performed ESR and pulse field magnetization measurements. A.K.B. analysed the experimental data with inputs from S.M.Y. S.K.S. and M.K. carried out the hybrid ED/DMRG calculations. A.K.B. and S.M.Y. wrote the manuscript, with contributions from S.K.S. and M.K. All authors discussed the data and its interpretation. S. M.Y. supervised the project.

**Competing interest:**

The authors declare no competing interests.

**Additional information:**

**Supplementary information**. The online version contains supplementary material available at ……………………..

Correspondence and requests for materials should be addressed to the corresponding author.

Reprints and permissions information is available at www.nature.com/reprints.

**Figures :**

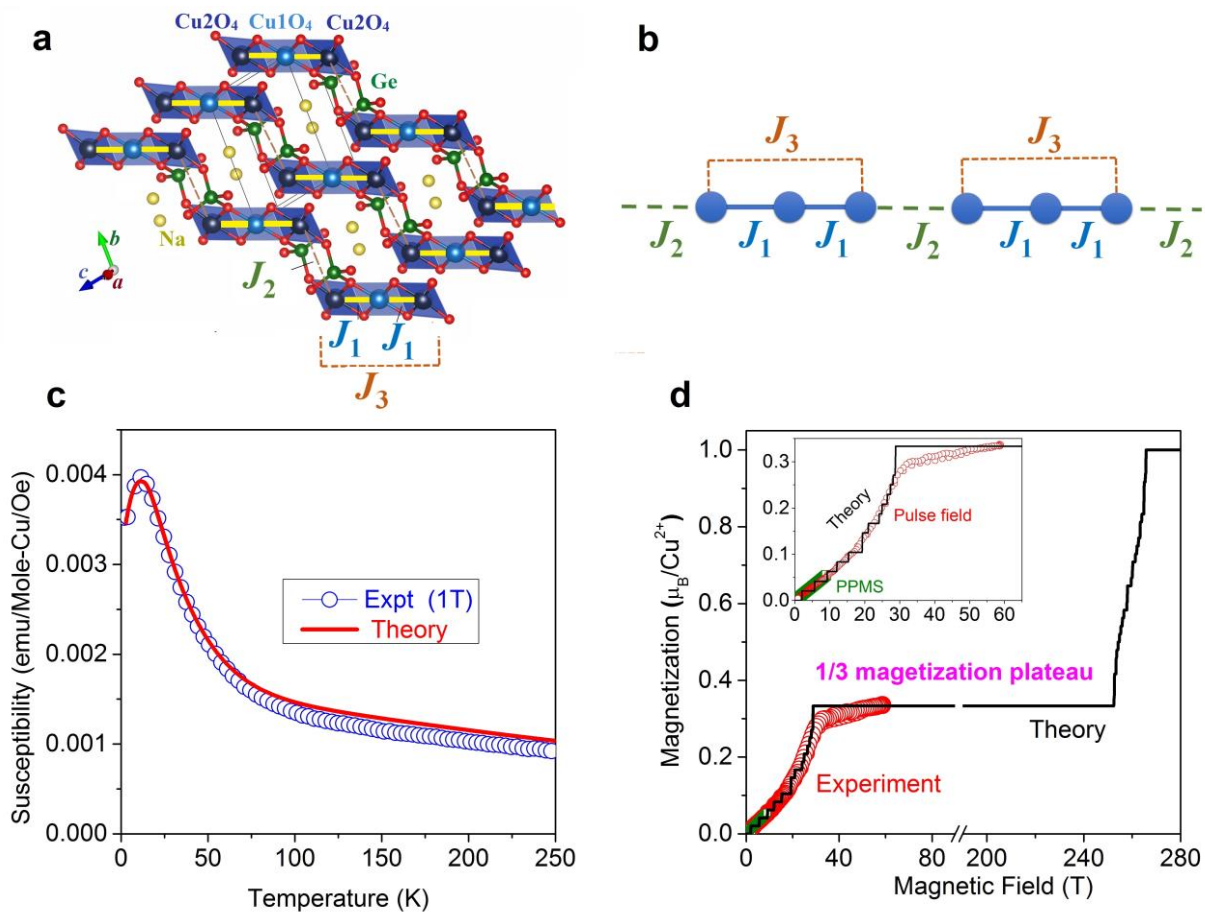

**Fig. 1| Crystal structure, magnetic model, magnetic susceptibility, and field dependent magnetization of $Na_2Cu_3Ge_4O_{12}$. a** The schematic spin-trimer structure of $Na_2Cu_3Ge_4O_{12}$. The spins S1, S2, and S3 are the three $Cu^{2+}$ spins within a trimer unit. The $J_1$, $J_2$, and $J_3$ denote the intratrimer, intertrimer and next-nearest neighbour intratrimer exchange couplings, respectively. **b** The schematics of 1D spin-chain with $J_1$, $J_2$ $(=\alpha J_1)$, and $J_3$ $(=\beta J_1)$. **c** The temperature dependent susceptibility ($\chi$ vs.$T$) curve (open circles) of $Na_2Cu_3Ge_4O_{12}$ measured under a magnetic field ($B$)= 1 Tesla. The calculated susceptibility curve as per the Hamiltonian (Eq.1) with the parameters $J_1$ = 235 K, $\alpha$ = 0.18, and $\beta$ = 0.18 is shown by the red solid line. The calculations were performed by a hybrid ED/DMRG method up to $N$=96. **d** The experimental isothermal magnetization ($M$) curve as a function of field $B$ (red circles) of $Na_2Cu_3Ge_4O_{12}$, measured at 3.0 K by static field up to 9 T [using a Physical Property Measurement System (PPMS)] and the pulse field up to 59 T (at HLD, Dresden), respectively. The magnetization plateau $M/M_s$ = 1/3 $\mu_B/Cu^{2+}$ above 28 Tesla is found. The calculated $M(B)$ curves of the model Hamiltonian in Eq.1 with the parameters $J_1$ = 235 K, $\alpha$ = 0.18, and $\beta$ = 0.18 for $N$ = 96 at $T$ = 0 is shown by the black line. The calculation indicate that the plateau state persists up to a magnetic field of 252.5 T. The small steps in the calculation at low field are artefacts arising from the finite size of the cluster.

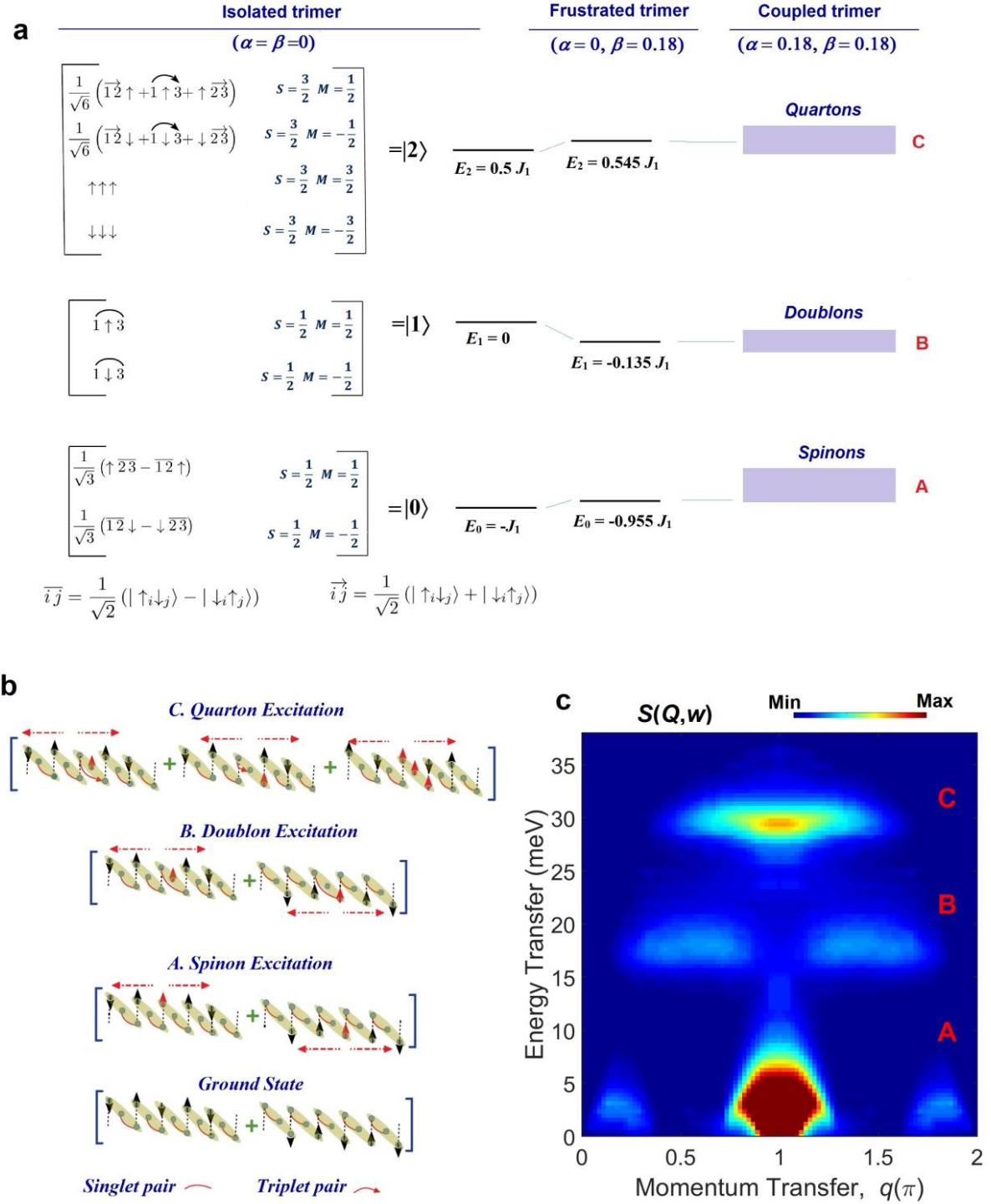

**Fig. 2| The energy level spectrum, wave functions, spin structures, and excitation spectrum of coupled spin-1/2 trimer-chain for the Hamiltonian of Na$_2$Cu$_3$Ge$_4$O$_{12}$. a** The level spectrum and corresponding wave functions of isolated trimer in presence of (left) only NN $J_1$ and (middle) both NN and NNN interactions [$J_1$, ($J_3=\beta J_1$ with $\beta$ = 0.18)]. The first column shows the wave functions where isolated up and down spins are represented by up and down arrows, respectively and the singlet and triplet pairing between sites is denoted by lines and arrows, respectively. The corresponding total spin quantum

number and magnetic quantum number are given by *S* and *M*, respectively. The last column shows the level spectrum for coupled trimers with $J_2$ (=$\alpha J_1$ with $\alpha$ = 0.18). The exchange interaction $J_2$ introduces dispersion for each of the energy levels where the lowest level (A), middle level (B), and highest level (C) excitations are composed of dispersive spinons, doublons, and quartons, respectively. **b** Possible spin configurations of the ground state, spinon, doublon and quarton excitations. The red arrows are the flipped spins with respected to the ground state. Red dashed arrows show the delocalization of spins throughout the system. The greenish oval represents the trimers and the intertrimer interactions are shown by black dashed lines. The up and down spins are depicted by uparrow and downarrow and color codes follow the text. **c** The dispersion relations and intensities [Dynamic spin structure factor $S(Q,w)$] for the spinon, doublon, and quarton excited states, over the full Brillouin zone along the chain, calculated by the DMRG (see Methods) for the Hamiltonian of $Na_2Cu_3Ge_4O_{12}$ ($J_1$ = 235 K, $\alpha$ = 0.18, $\beta$ = 0.18 and $g$ = 2.06).

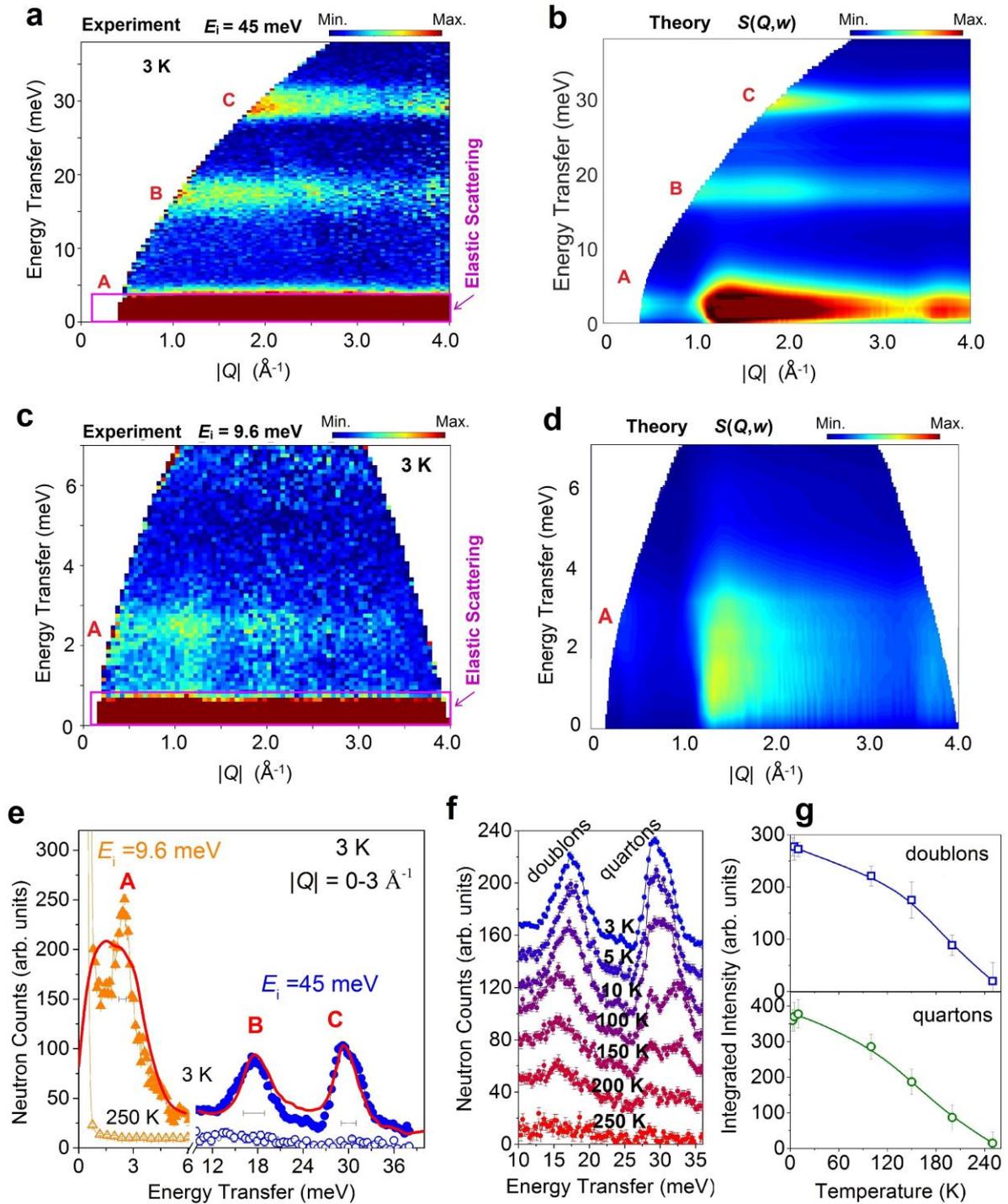

**Fig. 3| Dynamic structure factors of composite and spinon excitations in Na$_2$Cu$_3$Ge$_4$O$_{12}$ powders.**
**a,c** False-colour maps of the INS intensity measured at $T = 3$ K on the MERLIN spectrometer at ISIS facility, RAL, UK (in the 1D magnetic state above the $T_N$ = 2.25 K) using a fixed incident energy of $E_i$ = 45 and 9.6 meV, respectively (see Methods). The INS spectra have had an estimated phonon contribution subtracted (see Methods). The intensities are denoted by different colours, as indicated by the scales at each panel. The magenta boxes at the bottom of the each patterns represent the areas where the elastic scattering contributions are evident. **b,d** The excitation spectra for the spin-1/2 antiferromagnetic trimer-chain model for the Hamiltonian of $Na_2Cu_3Ge_4O_{12}$ ($J_1 = 235$ K, $\alpha = 0.18$, and $\beta = 0.18$) at $T = 0$, calculated using the hybrid ED/DMRG method. **e** Comparison of the scattering intensity with the theory. Energy cuts through the data (**a** and **c**) integrated over the |$Q$| range 0-3 Å$^{-1}$. Error bars indicate the standard deviation assuming Poisson counting statistics. The instrumental resolution at the peak energies are shown by the horizontal bars. The red solid curves are the calculated intensities. **f**, The temperature evolution of the doublons and quartons excitations which are found to be present up to a temperature~ 250 K. **g** The temperature dependent integrated intensity for the doublons and quartons excitations.

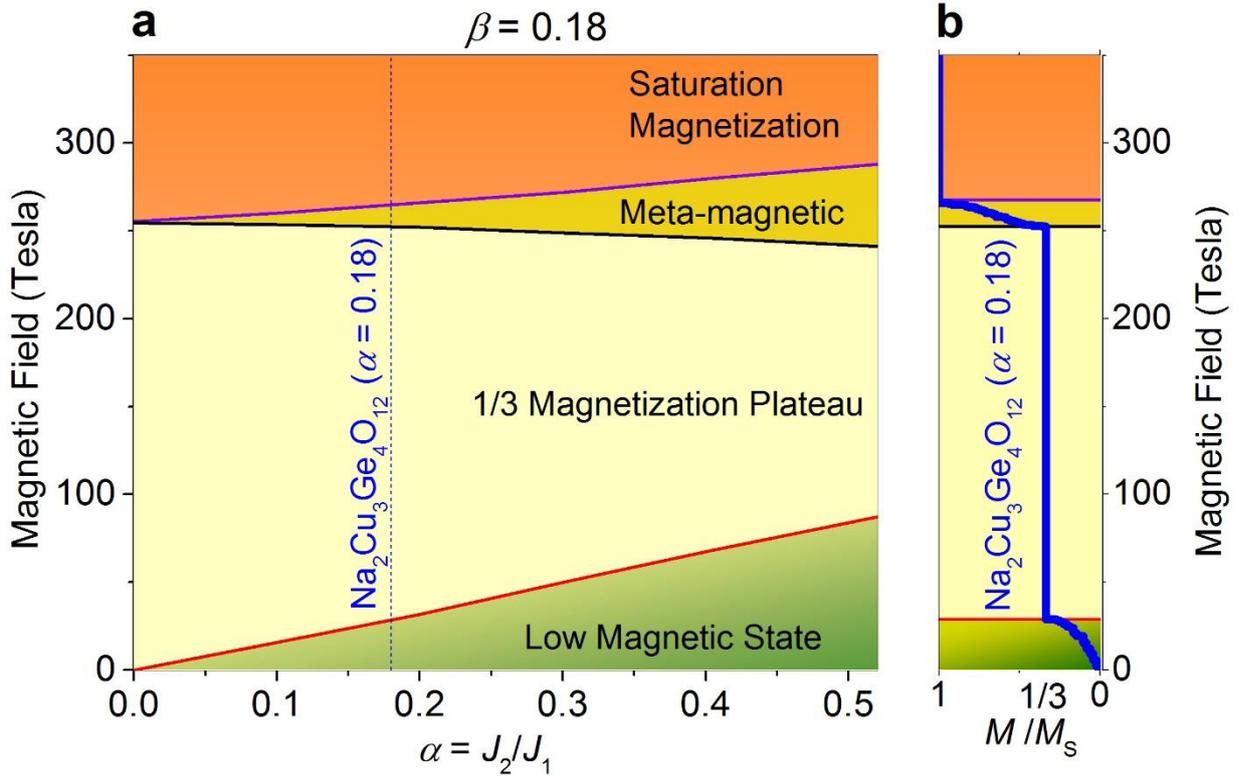

**Fig. 4| The magnetic phase diagram of Na$_2$Cu$_3$Ge$_4$O$_{12}$. a** The evolution of the magnetic field induced states as a function of the inter-trimer exchange coupling strength $\alpha = J_2/J_1$ (for $\beta = 0.18$). With the increasing $\alpha$ value, the 1/3 magnetization plateau state becomes narrower. The vertical dashed line represents the position of Na$_2$Cu$_3$Ge$_4$O$_{12}$. **b** The solid blue curve is the calculated magnetization curve for Na$_2$Cu$_3$Ge$_4$O$_{12}$ having $\alpha = 0.18$ [Fig. 1d] illustrating the different magnetic states, i.e., low magnetic state for $M/M_s < 1/3$, 1/3 magnetization plateau for $M/M_s = 1/3$, meta-magnetic state for $1/3 < M/M_s < 1$ and saturation magnetization for $M/M_s = 1$.

# Supplementary informations for

# Emergent many-body composite excitations of interacting spin-1/2 trimers

Anup Kumar Bera[1,2], S. M. Yusuf[1,2*], Sudip Kumar Saha[3], Manoranjan Kumar[3], David Voneshen[4,5], Yurii Skourski[6] and Sergei A. Zvyagin[6]

[1] *Solid State Physics Division, Bhabha Atomic Research Centre, Mumbai 40085, India*
[2] *Homi Bhabha National Institute, Anushaktinagar, Mumbai 400094, India*
[3] *S. N. Bose National Centre for Basic Sciences, Block JD, Sector III, Salt Lake, Kolkata 700106, India*
[4] *ISIS Facility, STFC Rutherford Appleton Laboratory, Harwell Oxford, Didcot OX11 0QX, UK*
[5] *Department of Physics, Royal Holloway University of London, Egham, TW20 0EX, UK*
[6] *Dresden High Magnetic Field Laboratory (HLD-EMFL), Helmholtz-Zentrum Dresden-Rossendorf, 01328 Dresden, Germany*

**e-mail: smyusuf@barc.gov.in**

## Supplementary Note 1: Crystal Structure

The crystal structure of $Na_2Cu_3Ge_4O_{12}$ has been investigated by the combined analysis of x-ray and neutron diffraction patterns at room temperature. The Rietveld analysis of the diffraction patterns (Supplementary Fig. 1) reveals that the compound crystallizes in the triclinic symmetry with group *P*-1 in agreement with that reported earlier. The lattice parameters are determined to be $a = 6.1824(3)$ Å, $b = 7.6912(3)$ Å, $c = 5.4721(2)$ Å, $\alpha = 102.388(2)°$, $\beta = 93.074(3)°$, and $\gamma = 87.579(3)°$ with unit cell volume $V = 253.65(2)$. The refined atomic positions, isotropic thermal parameters, and site occupation numbers are given in Supplementary Table-1.

In the present crystal structure, the magnetic $Cu^{2+}$ ions are distributed at two Wyckoff sites [Cu1(1*a*), and Cu2 (2*i*)]. The Na ions have single, Ge ions have two and O ions have six Wyckoff positions. The most prominent crystal structural feature of $Na_2Cu_3Ge_4O_{12}$ is the periodic arrays of $Cu_3O_8$ trimers formed by edge-sharing three $CuO_4$ square planes in a linear fashion with the intra-trimer copper distances of 3.048(5) Å, which is indicated by the nearly 180° dihedral angles. Within a given trimer, the central Cu1 atom is coordinated nearly square planar [coordination partners: 2 × O1 [Cu1-O1=1.915(6) Å] and 2 × O4 [Cu1-O4=2.049(7)]] and the two terminal Cu2 atoms are coordinated distorted square planar [coordination partners: O1 [2.001(9) Å] and O4 [1.957(7) Å], O5 [1.881(9) Å], and O6 [1.918(8) Å]]. The bridging angles between the square planar are Cu1–O1–Cu2=102.2(4) and Cu1–O4–Cu2= 99.0(4) (Supplementary Table-2). These trimers are embedded in an extended lattice, connected by corner sharing GeO$_4$ tetrahedra, exhibiting a pseudo-one-dimensional channel framework. The electro positive Na$^+$ cations reside in channels constructed by six corner-sharing polyhedral units (2 × CuO$_4$ and 4 ×GeO$_4$), as shown in Supplementary Fig. 2. In an isolated trimer, three Cu ions are coupled by intratrimer exchange interactions ($J_1$). Such trimers are coupled in a zigzag chain by intertrimer exchange interaction $J_2$ along the chain. Within a given trimer, the two edged Cu$^{2+}$ (Cu2) ions are coupled by the second nearest neighbour exchange interaction $J_3$. The details of the superexchange pathways are given in Supplementary Table-3. The Cu$_3$O$_8$ trimers chains are well separated by the nonmagnetic Na$^+$ and Ge$^{4+}$ ions resulting in a weak interchain interaction $J_{int}$.

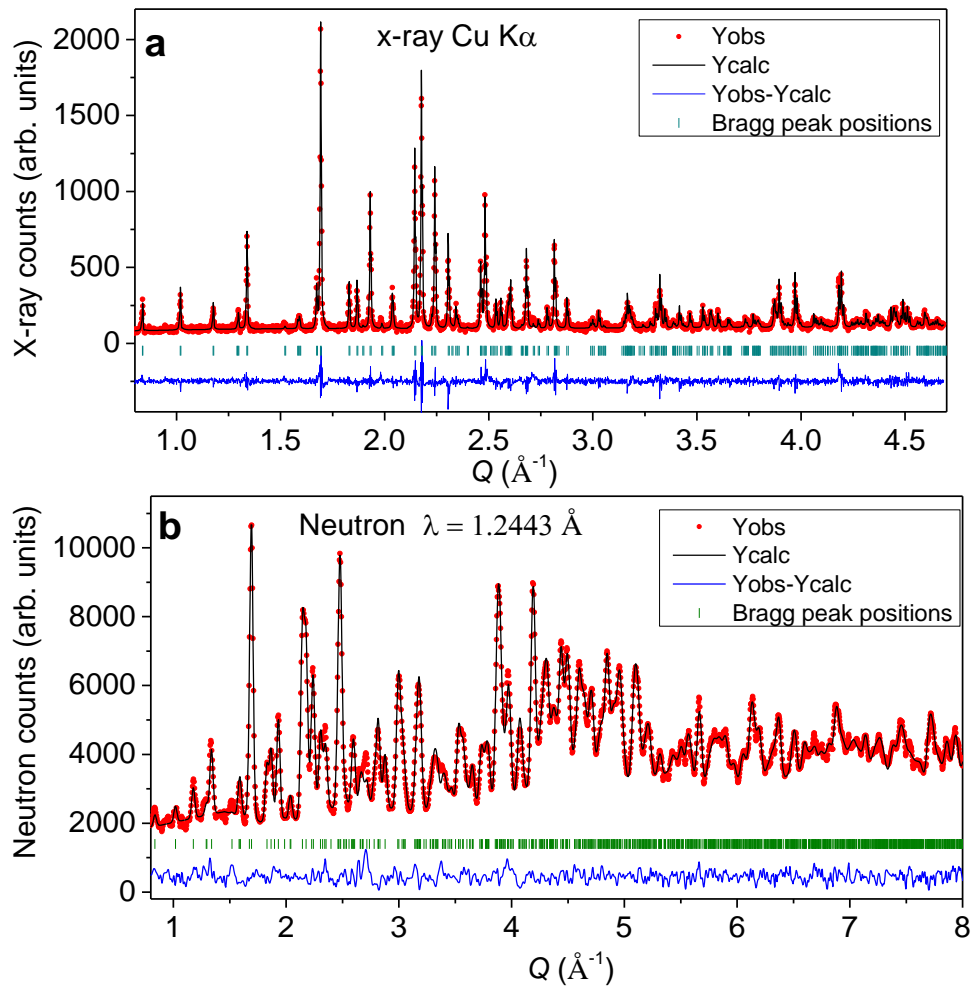

**Supplementary Fig. 1. a** and **b** The x-ray and neutron diffraction patterns, respectively, of Na$_2$Cu$_3$Ge$_4$O$_{12}$ at room temperature. Solid circles represent the experimental data points and the black line (through the data points) is calculated pattern by Reitveld method with the space group *P*-1.

**Supplementary Table 1.** The Rietveld refined atomic positions, isotropic thermal parameters, and site occupation numbers for $Na_2Cu_3Ge_4O_{12}$ at room temperature. Space group: $P$-1, $a$ = 6.1824(3), $b$=7.6912(3), $c$ = 5.4721(2), $\alpha$= 102.388(2), $\beta$ = 93.074(3), and $\gamma$= 87.579(3).

| Atom | Site | x/a | y/b | z/c | $10^2 \times B_{iso}$ (Å$^2$) | Occ. |
|---|---|---|---|---|---|---|
| Na | 2i | 0.2494(12) | 0.5955(13) | 0.9174(13) | 1.34(2) | 1.0 |
| Cu1 | 1a | 0 | 0 | 0 | 0.87(2) | 1.0 |
| Cu2 | 2i | 0.1632(14) | 0.8231(13) | 0.4873(12) | 0.70(1) | 1.0 |
| Ge1 | 2i | 0.2209(7) | 0.3676(6) | 0.3067(8) | 0.57(3) | 1.0 |
| Ge2 | 2i | 0.39881(8) | 0.18278(6) | 0.76148(9) | 0.64(2) | 1.0 |
| O1 | 2i | 0.0053(9) | 0.2246(8) | 0.2331(12) | 0.93(1) | 1.0 |
| O2 | 2i | 0.4021(9) | 0.3184(8) | 0.0644(11) | 1.04(3) | 1.0 |
| O3 | 2i | 0.3646(7) | 0.3493(7) | 0.5772(12) | 1.07(8) | 1.0 |
| O4 | 2i | 0.1749(11) | 0.0550(9) | 0.7197(12) | 1.20(3) | 1.0 |
| O5 | 2i | 0.6580(13) | 0.0965(7) | 0.7305(11) | 1.01(3) | 1.0 |
| O6 | 2i | 0.1273(9) | 0.5843(8) | 0.2996(10) | 1.05(4) | 1.0 |

**Supplementary Table 2.** The local crystal structural parameters for $Na_2Cu_3Ge_4O_{12}$; bond lengths, and bond angles at room temperature.

| Site | bond length (Å) | | | bond angle (°) | |
|---|---|---|---|---|---|
| Cu1 | Cu1–O1 | 1.915(6) | 2× | O1–Cu1–O1 | 180.0(7) |
|  | –O4 | 2.049(7) | 2× | O1–Cu1–O4 | 100.7(5) 2× |
|  |  |  |  |  | 79.3(5) 2× |
|  |  |  |  | O4–Cu1–O4 | 180.0(7) |
| Cu2 | Cu2–O1 | 2.001(9) |  | O1-Cu2-O4 | 79.5(5) |
|  | –O4 | 1.957(7) |  | O1-Cu2-O5 | 169.3(7) |
|  | –O5 | 1.881(9) |  | O1-Cu2-O6 | 91.5(5) |
|  | –O6 | 1.918(8) |  | O4-Cu2-O5 | 90.5(5) |
|  |  |  |  | O4-Cu2-O6 | 171.0(6) |
| Ge1 | Ge1–O1 | 1.740(8) |  | O1–Ge1–O2 | 109.0(6) |
|  | –O2 | 1.750(8) |  | O1–Ge1–O2 | 113.7(6) |

|       |         |          |             |          |
|-------|---------|----------|-------------|----------|
|       | –O3     | 1.714(8) | O1–Ge1–O6   | 109.0(6) |
|       | –O6     | 1.749(8) | O2–Ge1–O3   | 107.1(6) |
|       |         |          | O2–Ge1–O6   | 103.4(6) |
|       |         |          | O3–Ge1–O6   | 114.1(6) |
| Ge2   | Ge2–O2  | 1.759(7) | O2–Ge1–O3   | 100.2(6) |
|       | –O3     | 1.792(9) | O2–Ge1–O4   | 108.8(6) |
|       | –O4     | 1.708(9) | O2–Ge1–O5   | 104.2(6) |
|       | –O5     | 1.711(8) | O3–Ge1–O4   | 109.7(7) |
|       |         |          | O3–Ge1–O5   | 107.8(6) |
|       |         |          | O4–Ge1–O5   | 123.5(7) |

**Supplementary Table 3.** Possible pathways for the exchange interactions $J_1$, $J_2$ and $J_3$, respectively. The Cu...Cu direct distances, metal oxide (*M*–O) bond lengths and bond-angles for the exchange interactions $J_1$, $J_2$ and $J_3$ in $Na_2Cu_3Ge_4O_{12}$ at room temperature.

| Exchange interaction | Pathways | Cu...Cu direct distance (Å) | Bond lengths (Å) | Bond angles (deg.) |
|---|---|---|---|---|
| $J_1$ | Cu1–O–Cu2 | Cu1–Cu2= 3.048(5) | Cu1–O1= 1.915(6) Cu2–O1= 2.001(9) | Cu1–O1–Cu2=102.2(4) |
|  |  |  | Cu1–O4= 2.049(8) Cu2–O4= 1.957(8) | Cu1–O4–Cu2= 99.0(4) |
| $J_2$ | Cu1–O1–Ge1–O6–Cu2 (along *b* axis) | Cu1–Cu2= 6.420(5) | Cu1–O1= 1.915(6) Ge1–O1=1.740(8) Ge1–O6=1.749(9) Cu2–O6= 1.918(8) | Cu1–O1–Ge1=127.2(4) O1–Ge1–O6=109.0(4) Ge1–O6–Cu2=139.3(5) |
| $J_3$ | Cu2–O1–O4–Cu2 | Cu2–Cu2= 6.096(10) | Cu2–O1= 2.001(9) O1–O4=3.053(9) Cu2–O4= 1.957(8) | Cu2–O1–O4=143.3(4) O1–O4–Cu2=137.16(4) |
|  | Cu2–O1–Cu1–O4–Cu2 |  | Cu2–O1= 2.001(9) Cu1–O1=3.053(9) Cu1–O4= 1.957(8) Cu2–O4= 1.957(8) | Cu1–O1–Cu2=102.2(4) Cu1–O4–Cu2= 99.0(4) |

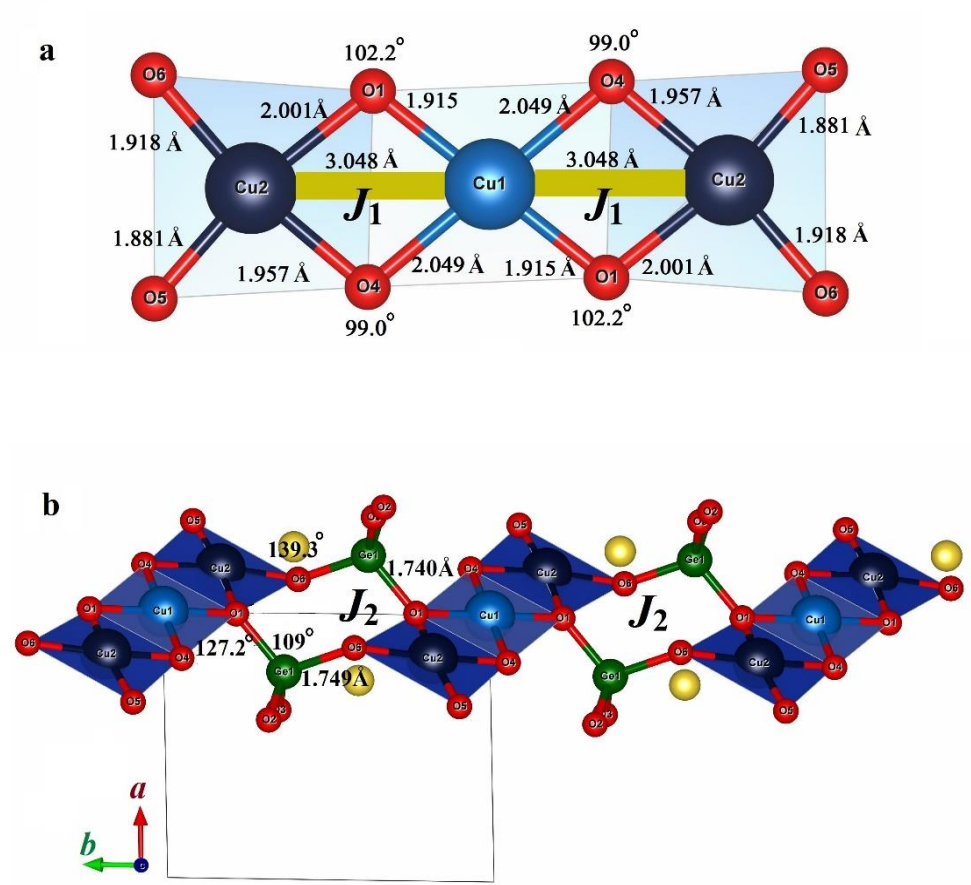

**Supplementary Fig. 2. a** The local crystal structure of an individual trimer unit of $Na_2Cu_3Ge_4O_{12}$. The experimentally measured values of the bond lengths, bond angles and the Cu-Cu distances at room temperature are shown. **b** The intertrimer connections along the crystallographic *b* axis through the $Ge_1O_4$ tetrahedra.

**Supplementary Note 2: Bulk magnetic properties:**

The temperature dependent susceptibility ($\chi$ vs $T$) curve measured under a magnetic field of $B = 1$ T is shown in Supplementary Fig. 3a. The $\chi(T)$ curve shows a broad maximum at $T_{max} \sim 11$ K due to the growth of the antiferromagnetic short-range order with decreasing $T$. With the further lowering of temperature, $\chi(T)$ curve shows an anomaly at $T \sim 2$ K (the inset of Supplementary Fig. 3a) revealing a magnetic long-range antiferromagnetic ordering in $Na_2Cu_3Ge_4O_{12}$ at $T_N \sim 2$ K. The magnetic long-range ordering below $T_N \sim 2$ K was reported by temperature dependent specific heat, nuclear magnetic resonance (NMR) and capacitance studies[22,23]. The $\chi T$ vs $T$ curve (Supplementary Fig. 3b) below $\sim 300$ K reveals a deviation from a constant value suggesting the deviation from the paramagnetic behaviour.

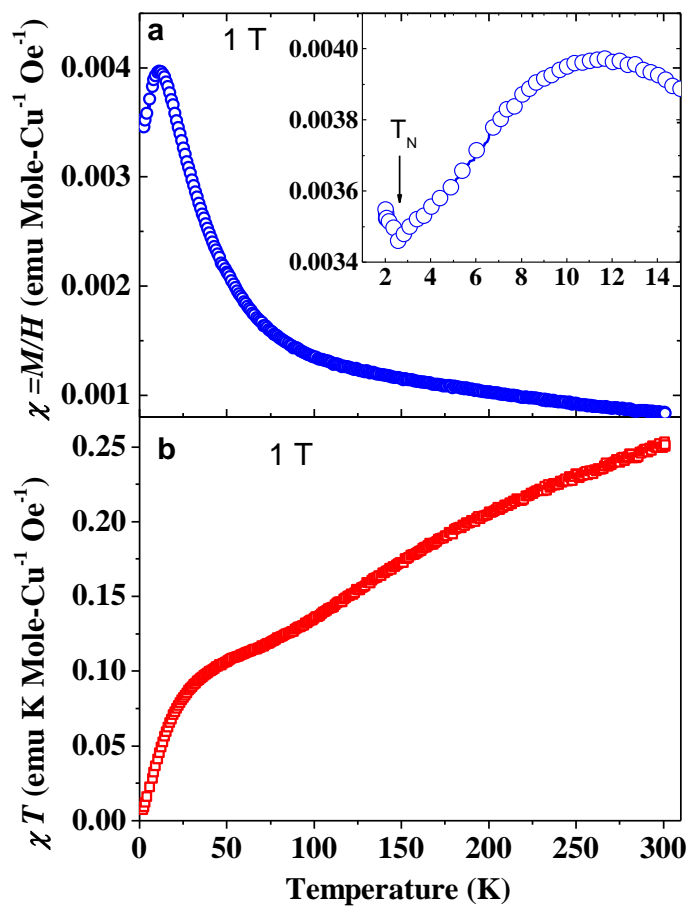

**Supplementary Fig. 3. a** The temperature dependent susceptibility [$\chi(T)$] curve measured on powder sample under a magnetic field $B = 1$ T. Inset highlights the low temperature region where a kink appears at the $T_N = 2$ K. **b** The $\chi T$ vs. $T$ curve revealing the deviation from the a constant value as expected for a paramagnetic state having Curie behaviour and indicates the presence of magnetic correlations up to about room temperature as compared to low magnetic ordering temperature $T_N \sim 2$. K.

**Supplementary Note 3: Electron Spin Resonance (ESR) spectroscopy:**

The ESR powder spectra measured at a frequency of 104 GHz in the 2 – 70 K temperature range are shown in Supplementary Fig. 4. The spectral weight has been found distributed approximately between 3.13 and 3.6 T, which for this frequency corresponds to the *g*-factors ranging from 2.36 to 2.06, respectively. The spectra exhibit a tiny but noticeable broadening upon approaching the transition into the magnetically ordered state at $T_N$ =2 K (compare, e.g., the spectra at 10 and 2 K).

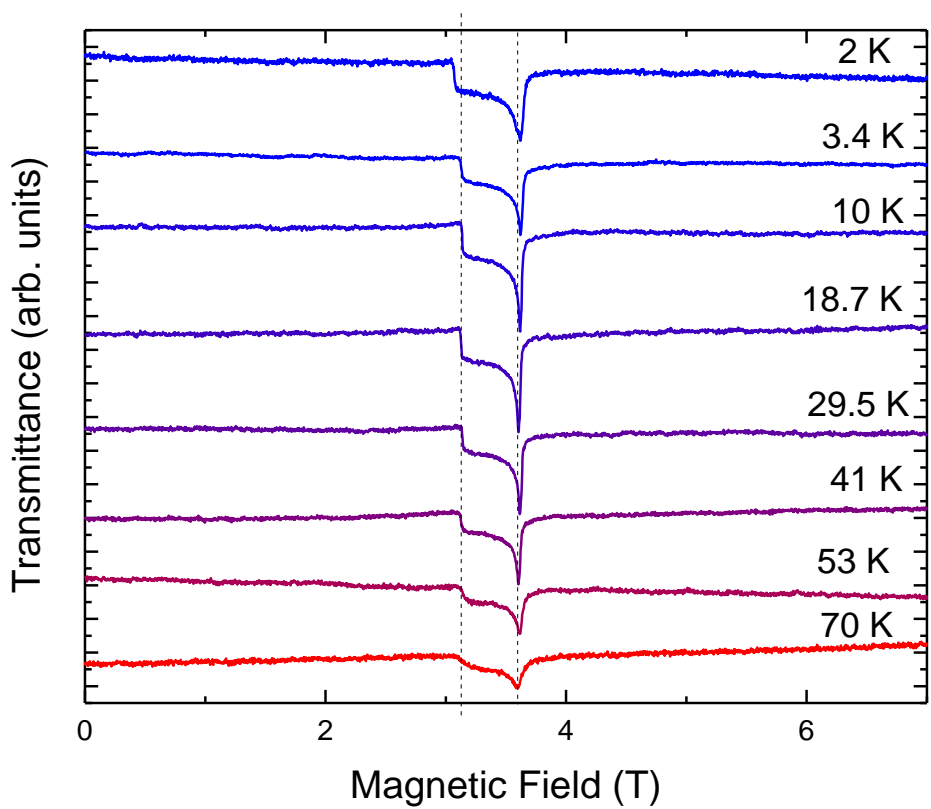

**Supplementary Fig. 4.** Examples of ESR spectra in $Na_2Cu_3Ge_4O_{12}$, measured at a frequency of 104 GHz at different tempertaures (the spectra are offset for clartity). The dash lines correspond to the *g*-factors 2.36 and 2.06.

**Supplementary Note 4: Quantum Entanglement Parameter:**

Quantum entanglement is a physical phenomenon that occurs when a group of particles interact among each other such a way that the quantum state of individual particle cannot be described independently of the state of the others, including when the particles are separated by a large distance. Entanglement is a primary feature of quantum mechanics lacking in classical mechanics. Quantum entanglement, the most prominent example of quantum correlations, serves as the fundamental resource in several quantum information processing tasks such as quantum computation and communication[41]. To measure quantum entanglement as a physical quantity, by using the Hilbert-Schmidt norm[42], Del Cima *et al.*[18], formulated a general method for spin-1/2 antiferromagnetic trimer. The weak inter trimer exchange interaction ($\alpha \ll 1$) are neglected for simplicity. The quantum entanglement measure $E(J,T)$ as a function of the temperature ($T$) and the exchange coupling ($J$) is defined as

$$E(J,T) = E_0 max\left[0, \left(2\left|\frac{3}{8}\frac{1+e^{\frac{j}{k_BT}}+10e^{\frac{3j}{2k_BT}}}{1+e^{\frac{j}{k_BT}}+2e^{\frac{3j}{2k_BT}}}-1\right| + \frac{1}{2} - \frac{3}{8}\frac{1+e^{\frac{j}{k_BT}}+10e^{\frac{3j}{2k_BT}}}{1+e^{\frac{j}{k_BT}}+2e^{\frac{3j}{2k_BT}}}\right)\right] \quad (1)$$

with $\lim_{T \to 0} E(J,T) = \frac{11}{8}E_0 = \frac{11}{32}$

The entanglement critical temperature ($T_c$) is defined as the temperature where the quantum entanglement measure, $E(J,T)|_{T=T_c} \equiv 0$ vanishes. Thus, for temperatures below the critical (decoherence) temperature ($T < T_c$), the system is entangled, whereas, for temperatures above it, $T > T_c$, the system experiences quantum decoherence, it assumes a separable state. We have estimated the quantum entanglement measure $E(J,T)$ for $Na_2Cu_3Ge_4O_{12}$ following the above equation. The temperature dependence of the $E(J,T)$ is shown in Supplementary Fig. 5 which reveal a very high $T_c$ of ~ 310 K as the exchange coupling protects the system from decoherence as temperature increases. Such a high decoherence temperature makes $Na_2Cu_3Ge_4O_{12}$ as a suitable candidate for practical device applications.

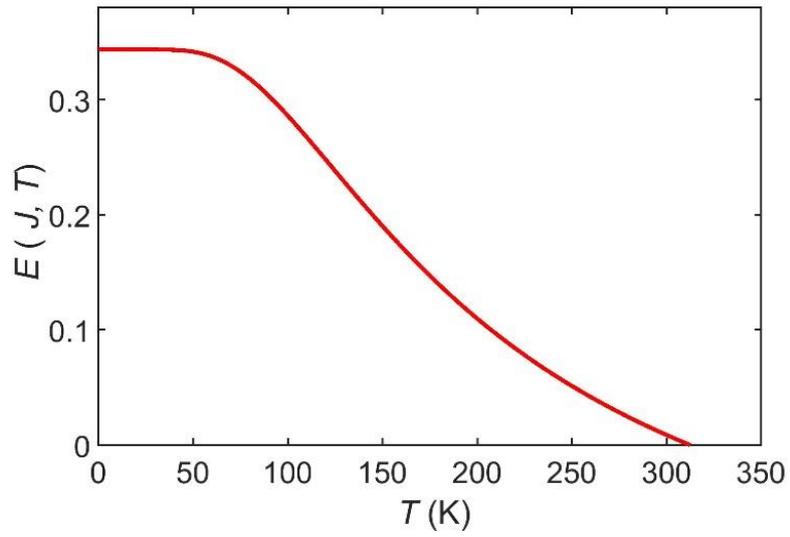

**Supplementary Fig. 5.** The quantum entanglement measure, $E(J,T)$, for $Na_2Cu_3Ge_4O_{12}$, with $J/k_B$ = −235 K.

**Supplementary Note 5: Phase diagram with the frustrated interaction $J_3$ ($\beta$-$J_1$):**

The phase diagram of trimer model in Eq. 1 in $\beta$-$H$ parameter space is shown for $\alpha=0$ and 0.18 at $J_1 = 235$ K in Supplementary Figs. 6a and 6b, respectively. For $\alpha = 0$ and finite $\beta$, the system behaves as isolated triangles with $\beta$ as exchange interaction at the base of the triangle. The ground state is a doublet $M = 1/3$ state and has a finite gap which corresponds to gap between doublet and quartet states, therefore it requires large field to achieve the higher magnetization as shown in Supplementary Fig.6a. For $\alpha = 0.18$ and $\beta \leq 1.5$, each trimer has effective spin 1/2 which are coupled antiferromagnetically and forms a singlet ground state. It requires finite field to reach the $M = 1/3$ plateau phase and at higher values of $H$ the system accesses metamagnetic phase and saturated magnetization. Large $\beta$ (~1 or >1) favours the weakly coupled triangles configuration where base of each triangle has strong AFM coupling forming a strong dimer singlet. For $1.1 < \beta < 1.5$, spins at the apex of a triangle are coupled effectively with weak AFM exchange and the ground state is still in the singlet state and it requires small $H$ to achieve $M = 1/3$ plateau state. For $\beta \geq 1.5$, the apex spins are coupled ferromagnetically resulting $M = 1/3$ plateau state as a ground state and this state has large energy gap. At $\beta \sim 1$ the width of the metamagnetic state increases due to formation of strong singlet dimers at the base of the triangles and it is proportional to the energy required to break the singlet dimers.

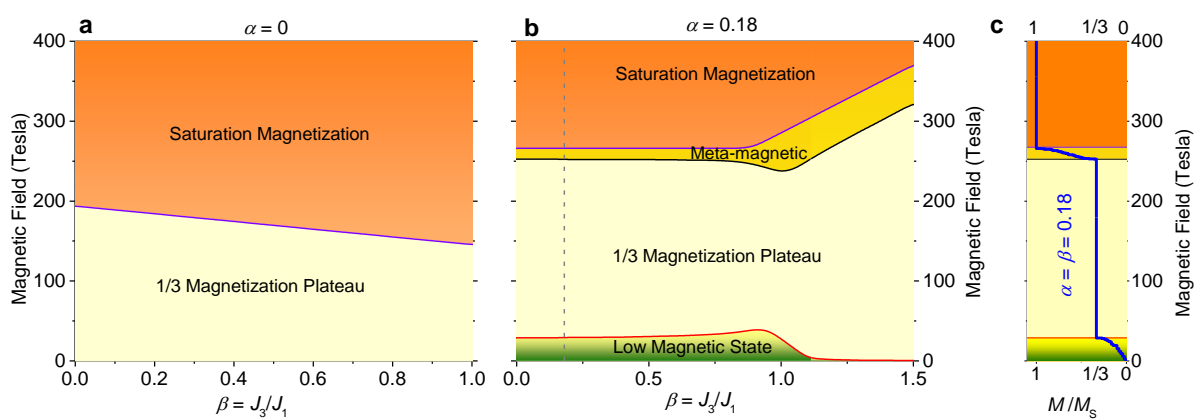

**Supplementary Fig. 6.** The quantum phase diagram of the trimer model in Eq. 1 in $\beta$-$H$ parameter space at $J_1 = 235$ K for **a** $\alpha=0$ and **b** $\alpha=0.18$. **c** the definition of the magnetic states with a representative magnetization curve for $Na_2Cu_3Ge_4O_{12}$ ($\alpha=\beta=0.18$), i.e., low magnetic state for $M/M_s < 1/3$, 1/3 magnetization plateau for $M/M_s = 1/3$, meta-magnetic state for $1/3 < M/M_s < 1$ and saturation magnetization for $M/M_s =1$.

**Supplementary Note 6: Comparison of simulations of neutron structure factors:**

The DMRG with the correction vector method is a well-established numerical technique to calculate the dynamical structure factor $S(q,\omega)$ [40,43-45] and other dynamical properties [38,39] for various model 1D spin systems. To compare the DMRG calculated $S(q,\omega)$ with the results derived from all the three calculation methods (ED, ED with truncated Hilbert space, QMC-SAC calculations) reported in Refs. [20,26], we use $N=48$ spins with broadening factor ($\eta=0.07$). We calculate $S(q,\omega)$ at $\alpha=0.2$ and $\beta=0$ and compare the results for $g=0.2$ in Refs. [20,26], i.e, (i) ED with truncated Hilbert space with $N = 24$ spins (Fig. 8b of Ref. 12), (ii) QMC-SAC calculation with $N = 192$ spins, and $\eta=0.05$ (Fig 2b of Ref. 12) and, (iii) ED for $N=30$ spins [Fig. 2(b) of Ref. [26]]. It is evident that the calculated $S(q,\omega)$ in the present study (Supplementary Fig. 7) revealing the signatures of spinon, doublon and quarton excitation modes are in good agreement with the reported values (Refs. [20,26]). The energy modes of the doublon and quarton match well with the calculated results using all the three methods. For the spinon modes, the lower energy region matches well with all the three methods, whereas, the higher energy region matches better with the ED calculation. The QMC-SAC results may be reproduced using a larger system size in the present calculation involving the DMRG with correction vector method.

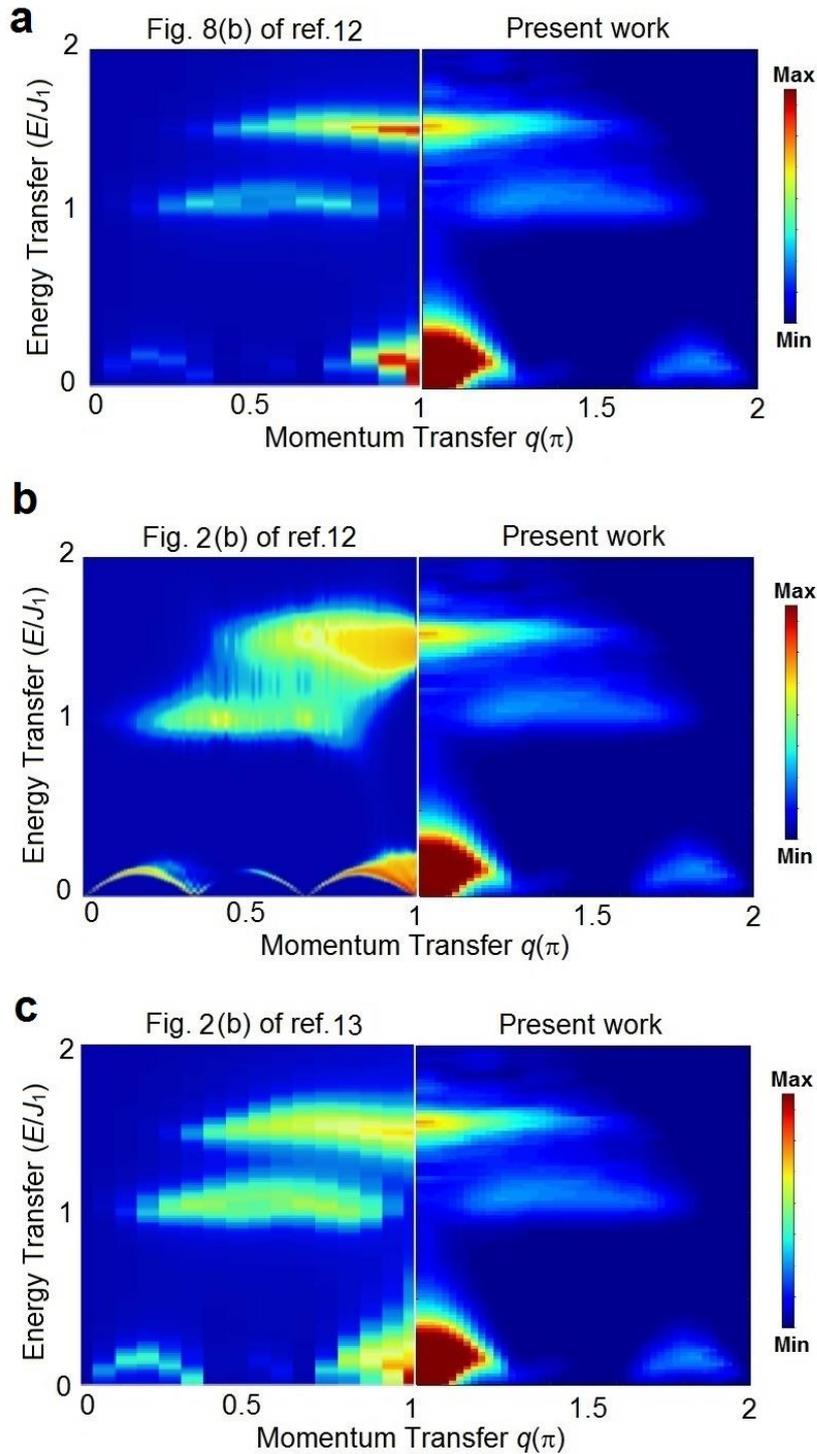

**Supplementary Fig. 7.** Comparison of the calculated dynamic structure factors calculated by **a** ED with truncated Hilbert space [Fig. 8(b) of Ref. [20]] and **b** Quantum Monte Carlo (QMC) by applying a variant of the stochastic analytic continuation [SAC] [Fig. 2(b) of Ref. [20]] and **c** ED [Fig 2(b) of Ref. [26]] methods with the present calculation by ED/DMRG method.